%% file: paper.tex
\documentclass[journal]{IEEEtran}

\usepackage[T1]{fontenc}
\usepackage[utf8]{inputenc}
\usepackage{textcomp}
\usepackage{graphicx} 
\usepackage{amsmath} 
\usepackage{amssymb}  
\usepackage{mathtools}
\mathtoolsset{showonlyrefs}
\usepackage{subcaption}
\usepackage{flushend}
\usepackage{url}
\usepackage{color}
\newcommand{\code}[1]{\texttt{#1}}

\usepackage{tikz}
\usetikzlibrary{calc}
\usetikzlibrary{shapes.multipart, arrows.meta, positioning, decorations.markings, shapes.geometric}

\tikzset{
  sysmlblock/.style={
    draw,
    rectangle split,
    rectangle split parts=2,
    rectangle split part align={center,left},
    text width=6cm,
    minimum height=1.5cm,
    node distance=1.5cm,
    font=\sffamily,
    align=left,
    fill=gray!10
  },
  arrow/.style={
    -{Stealth},
    thick,
    shorten >=2pt,
    shorten <=2pt
  },
  composition/.style={
    -{Stealth},
    thick,
    shorten >=2pt,
    shorten <=2pt,
    postaction={decorate,decoration={markings,mark=at position 0.5 with {\fill (0,0) -- (2pt,2pt) -- (0,4pt) -- (-2pt,2pt) -- cycle;}}}
  }
}

\tikzset{
  block/.style={
    draw,
    rectangle,
    minimum width=2cm,
    minimum height=1cm,
    fill=gray!10,
    font=\sffamily\small,
    align=center
  },
  port/.style={
    draw,
    rectangle,
    minimum width=1.2cm,
    minimum height=0.5cm,
    font=\sffamily\scriptsize,
    fill=white
  },
  arrow/.style={
    -{Stealth},
    thick
  }
}

\IEEEoverridecommandlockouts

\mathtoolsset{showonlyrefs}

\begin{document}

% author names 
\author{Sebastian~Dingler$^1$ (orcid: 0000-0002-0162-8428)\thanks{$^1$Daimler Truck AG, Stuttgart, Germany\\\texttt{sebastian.dingler@daimlertruck.com}}, Philip~Rehkop$^2$,\\
Florian~Mayer$^2$ (orcid: 0009-0002-5303-8830), Ralf~Münzenberger$^2$ \thanks{$^2$INCHRON AG, Erlangen, Germany\\\texttt{$\{$philip.rehkop, florian.mayer,\\ ralf.muenzenberger$\}$@inchron.com}}}

% Schlachtplan
% Am Anfang des Papers erwähnen, dass Regulatorik auch fordert dass Dokumentation beweist 
% Erwähnen dass es hier um Systems Engineering geht also vor der Verfügbarkeit einer Komponente also um Shift Left geht
% Swimmlane diagram mit der Allokation der Funktionen auf die ECUs
% Funktion für das ausgeben der Warnung hinzufügen
% Darstellung der 3 Phasen: 1. Warnung, 2. Abbremsen, 3. Bremsen
% Herleiten was ist die Sensormindestreichweite um noch die Notbremsung zu machen
% In der Simulation überprüfen, wie mit Verarbeitungszeiten umgegangen wird
% Als Optimierung darstellen entweder dass die Frequenz, die Sensordetektionrate oder Sensorreichweite verbessert werden kann
% Regulatorik mit 0.6 seconds zitieren
% Say that the paper demonstrate that event-chains analysis is capable of modeling random effects like the sensors probability of detection

\title{Event-Chain Analysis for Automated Driving and ADAS Systems: Ensuring Safety and Meeting Regulatory Timing Requirements}

\maketitle

\begin{abstract}
Automated Driving Systems (ADS), including Advanced Driver Assistance Systems (ADAS), must fulfill not only high functional expectations but also stringent timing constraints mandated by international regulations and standards. 
Regulatory frameworks such as UN regulations, NCAP standards, ISO norms, and NHTSA guidelines impose strict bounds on system reaction times to ensure safe vehicle operation.

This paper presents a structured, \emph{White-Box} methodology based on \emph{Event-Chain Modeling} to address these timing challenges. 
Unlike \emph{Black-Box} approaches, \emph{Event-Chain Analysis} offers transparent insights into the timing behavior of each functional component—from perception and planning to actuation and human interaction.
This perspective is also aligned with multiple regulations, which require that homologation dossiers provide evidence that the chosen system architecture is suitable to ensure compliance with the specified requirements.
Our methodology enables the derivation, modeling, and validation of end-to-end timing constraints at the architectural level and facilitates early verification through simulation.
    
Through a detailed case study, we demonstrate how this \emph{Event-Chain}-centric approach enhances regulatory compliance, optimizes system design, and supports model-based safety analysis techniques, with results showing early identification of compliance issues, systematic parameter optimization, and quantitative evidence generation through probabilistic analysis.
\end{abstract}

%\begin{IEEEkeywords}
%keywords, temperature, xxxx equation, etc.
%\end{IEEEkeywords}

\section{Introduction}
% Here we have the typical use of a "W" for an initial drop letter
% and "RITE" in caps to complete the first word.

\IEEEPARstart{T}{he} development of Advanced Driver Assistance Systems (ADAS) and Automated Driving Systems (ADS) is increasingly shaped by regulatory requirements. 
These regulations extend across various domains—including safety, cyber-security, and artificial intelligence—and impose rigorous constraints throughout the engineering lifecycle. 
In particular, compliance with regulations such as UN regulations, NCAP standards, ISO norms, and NHTSA guidelines has become a critical aspect of system design and validation.

Non-compliance with such regulations can result in consequences: high financial penalties, product recalls, legal disputes, reputational damage, production delays, sales bans, and disruptions across the supply chain. 
From a systems engineering perspective, the most effective strategy to mitigate these risks is to manage them proactively and early in the development cycle. 
This ``shift-left'' approach involves predictive modeling and simulation to identify and resolve issues before they manifest in late-stage integration or vehicle testing.

Given the increasing complexity of automotive systems—spanning heterogeneous ECUs, cross-domain functions, and evolving software-defined architectures—we use the INCHRON-proposed \emph{Event-Chain Analysis} \cite{MayerKraft2023}\cite{Heckmann.2021} as a representative methodology for verifying system behavior with a particular focus on timing requirements, which are both explicitly defined and implicitly derived from legal and regulatory standards. 
This choice is further motivated by regulatory expectations for auditable documentation---for example, ``the manufacturer shall demonstrate, via appropriate documentation, that the system is capable of reliably delivering the required performances''---which \emph{Event-Chain Models} naturally provide as structured, verifiable evidence.
\emph{Event-Chains} offer a \emph{formal, verifiable} means of specifying system behavior from perception to actuation, enabling \emph{White-Box} simulation and tracing with clear regulatory links.

Previous work~\cite{MayerKraft2023} has demonstrated \emph{Event-Chain}-based approaches for managing zonal vehicle architecture complexity, introducing an \emph{Event-Chain} systems-architecture perspective to (i) structure end-to-end functions across zones and domains, (ii) expose timing-relevant interfaces between logical functions and platform services, and (iii) support design decisions on allocation and communication in a software-defined vehicle context.
Their work highlights how \emph{Event-Chain} viewpoints clarify responsibility splits between feature logic and base-platform services and improve cross-domain coordination. 
Complementary to this, \cite{Heckmann.2021} reports an industrial \emph{Event-Chain}-centric architecture design workflow for driver assistance systems that allocates chains to platform elements, budgets end-to-end timing, and uses tool-supported architectural simulation to validate feasibility.
In contrast to \cite{MayerKraft2023}, we add a regulation-to-design ontology linking legal clauses to architecture elements, formalize warning/braking trigger rules, incorporate probabilistic sensor detection, and demonstrate simulation-backed verification of regulatory timing early in design.

While existing methodologies—such as certification-compliant \emph{Event-Chain Modeling} approaches—focus heavily on traceability and information management, the methodology presented here is distinct in its emphasis on \emph{timing-aware} behavioral validation \cite{Graesler.2023}. 
This allows engineers not only to demonstrate logical traceability but also to validate that the system meets real-time operational constraints defined by regulatory standards.

% ##############################################################
\input{sec02_regulatory}

% ##############################################################
\input{sec03_ec_methodology}

% ##############################################################
\input{sec04_case_study}

% ##############################################################
\section{Conclusion}

This paper has presented a comprehensive \emph{Event-Chain} methodology for ensuring regulatory compliance of Advanced Driver Assistance Systems (ADAS) and Automated Driving Systems (ADS), with particular emphasis on timing requirements. The work addresses a critical gap in automotive systems engineering by providing a structured, \emph{White-Box} approach to timing validation that bridges the gap between regulatory text and technical implementation.

The methodology presented in this paper has been applied and evaluated within the Daimler Truck development environment, with a particular focus on regulatory timing requirements. It has proven to be a strong candidate for bridging methodological rigor with industrial pragmatism, showing clear potential for adoption in future vehicle programs.

The key contributions are threefold: (1) a regulation-to-design ontology that systematically translates legal clauses into technical requirements, (2) an \emph{Event-Chain} methodology enabling \emph{White-Box}, timing-aware analysis across perception, decision, and actuation, and (3) demonstration through an AEB case study showing how probabilistic sensor effects and braking dynamics can be embedded into executable models for compliance evaluation.

The methodology offers regulatory alignment through direct reflection of certification requirements, comprehensive traceability, and cross-domain applicability. The case study demonstrates that baseline configurations can underperform on timing requirements and that targeted adjustments can restore compliance, while quantifying relationships between timing budgets, sensor detection ranges, and actuation dynamics.

Implementation was supported by INCHRON's chromSIM simulation platform, which provides \emph{Event-Chain Modeling} and Monte Carlo simulation capabilities suitable for regulatory compliance verification in safety-critical automotive applications.

From an OEM perspective, \emph{Event-Chain Analysis} holds potential to support regulatory compliance efficiently, enable early detection of non-compliance, and balance methodological rigor with practical feasibility. From our perspective, the approach is therefore considered a highly promising methodology to reconcile certification-oriented engineering with industrial feasibility.

\section*{Transparency Statement}

Transparency on generative AI use: We employed ChatGPT (OpenAI, GPT-4/5, accessed August 2025) to support content development (brainstorming alternatives, generating outlines, and rephrasing text for clarity). The tool was not used to conduct analyses, create results, or verify factual claims. All factual statements, equations, figures, tables, and citations were produced and validated by the authors, who accept full responsibility for the content.

\bibliographystyle{IEEEtran}
\bibliography{bibliography}
\end{document}

%% file: sec02_regulatory.tex
\section{Regulatory Background and Timing Constraints}
The automotive industry, particularly in Europe, is among the most heavily regulated sectors, with over 150 EU regulations and 30 directives governing safety, environmental impact, and operational standards \cite{ACEA.2023}.
Within this landscape, Advanced Driver Assistance Systems (ADAS) and Automated Driving Systems (ADS) face not only functional requirements but also strict timing constraints that determine how quickly they must react to driving situations.
Timing requirements originate from two primary sources:
\begin{enumerate}
    \item \textbf{Explicit timing constraints in regulations and standards}, which specify concrete thresholds (e.g., minimum warning times or maximum allowable delays).
    \item \textbf{Implicit timing constraints derived from regulatory requirements}, such as safety margins and operational boundaries, which translate into time-critical conditions the system must respect.
\end{enumerate}

Explicit examples include:

\begin{itemize}
    \item \textbf{UN Regulation No. 152 (AEBS)} mandates that a forward collision warning must be issued at least \textbf{0.8 seconds} before autonomous braking is initiated.
    
    \item \textbf{UN Regulation No. 13} outlines general braking requirements, including timing constraints.
    
    \item \textbf{UN Regulation No. 79} addresses steering and lane-keeping systems, including automated lateral control with constraints on system activation timing.
    
    \item \textbf{ISO 15622} (Adaptive Cruise Control) specifies acceptable system response times up to \textbf{2.0 seconds} for longitudinal deceleration.
    
    \item \textbf{NHTSA Guidelines} (USA) emphasize \textbf{minimum response times}, fallback strategies, and driver engagement protocols.
\end{itemize}

Implicit timing boundaries follow directly from vehicle dynamics. For example, a regulation stating “the vehicle shall brake to prevent a collision” implicitly requires braking within the time window dictated by speed and deceleration. At highway speeds, every $100$ms of additional system latency translates into several extra meters of travel—imposing hidden timing demands on sensors, processing, and actuation.

Thus, timing is not merely a regulatory box-check: it directly impacts sensor specifications, system feasibility, and ultimately vehicle cost. To make these requirements actionable, we apply a structured ontology that links legal text to technical models. This enables traceability, reduces ambiguity, and allows formal verification using \emph{Event-Chains}.

Finally, while regulations are written as binary pass/fail conditions, real systems inevitably operate with probabilistic variation due to sensors, actuators, and environmental factors. Probabilistic modeling does not contradict regulatory intent; instead, it reframes compliance as a demonstrable, measurable quality level rather than an unrealistic absolutism.

%% file: sec03_ec_methodology.tex
\section{Event-Chain Engineering to Ensure Regulatory Timing Requirements}

Ensuring regulatory compliance requires a precise and traceable representation of the system's dynamic behavior.
As introduced in previous work on \emph{Event-Chain Analysis} \cite{MayerKraft2023, Heckmann.2021}, this methodology offers a structured, \emph{White-Box} approach to model, trace, and validate the timing behavior.

Starting with regulatory texts, this section introduces the methodology, which integrates formal \emph{Event-Chain} modeling with systems engineering practices to meet certification standards effectively.

First we provide definitions of \emph{Event-Chains} and related terms. Afterwards we show how \emph{Event-Chain} models can be derived from regulatory texts. For this an ontology with multiple formalization steps is provided, that also illustrates the relations to different views on the system. The models are translated from the initial \emph{Black-Box View} into \emph{White-Box View}. Finally the models can be used for analyses.

\subsection{Modeling Event-Chains as a Basis for Temporal System Evaluation}

\emph{Event-Chain Engineering} is a model-based approach for evaluating systems based on the sequences of discrete events -- referred to as \emph{Event-Chains} -- that they emit. This methodology is conceptually aligned with \emph{Discrete Event Systems} (DES), which characterize system behavior as being entirely governed by the occurrence of discrete, asynchronous events that trigger state transitions \cite[p. 1311]{silva_modeling_2021}.

Each \emph{Event-Chain Model} defines a subset of event sequences and the causal dependencies between their constituent elements, its \emph{Event-Chain Steps}. An \emph{Event-Chain Step} comprises a named event emitted by the system, along with zero or more references to other \emph{Event-Chain Steps} on which it causally depends. 
While not all steps need to be directly connected, every \emph{Event-Chain Step} must be transitively dependent on exactly one common start event -- the unique \emph{Event-Chain Step} within the model that has no incoming dependency links.

Timing requirements (such as latency, synchronization, periodicity, or data age) are specified between individual \emph{Event-Chain Steps}.

\emph{Event-Chain Models} may exhibit dependencies among each other, as many behavioral patterns require certain system conditions to be met prior to their initiation. For instance, initiating an emergency brake action assumes that up-to-date information about the vehicle's current motion and position is already available. Modeling the acquisition of this data as part of the same \emph{Event-Chain} would be misleading, since it may already be ongoing or completed when the start event occurs. To address this, such preparatory behaviors are modeled as separate \emph{Event-Chain Models}, and explicit inter-model dependencies are introduced. These dependencies, however, are outside the scope of this work. 

The evaluation of a system with respect to its \emph{Event-Chain Models} is referred to as \emph{Event-Chain Analysis}. This includes, among other aspects, the verification of timing requirements that are formulated \emph{with reference to} the \emph{Event-Chains} -- specifically, the timing relationships between individual \emph{Event-Chain Steps}. \emph{Event-Chain Analysis} can be performed either using a timing model of the system -- via simulation or formal analysis -- or by evaluating execution traces recorded from the target system.

\input{sec03.01_requirements}

\input{sec03.02_ec_models}

\input{sec03.03_analysis}

%% file: sec03.01_requirements.tex
\subsection{Deriving Event-Chain Models from Regulatory Texts}

\begin{figure*}[h!]
    \centering
    \includegraphics[width=0.8\textwidth]{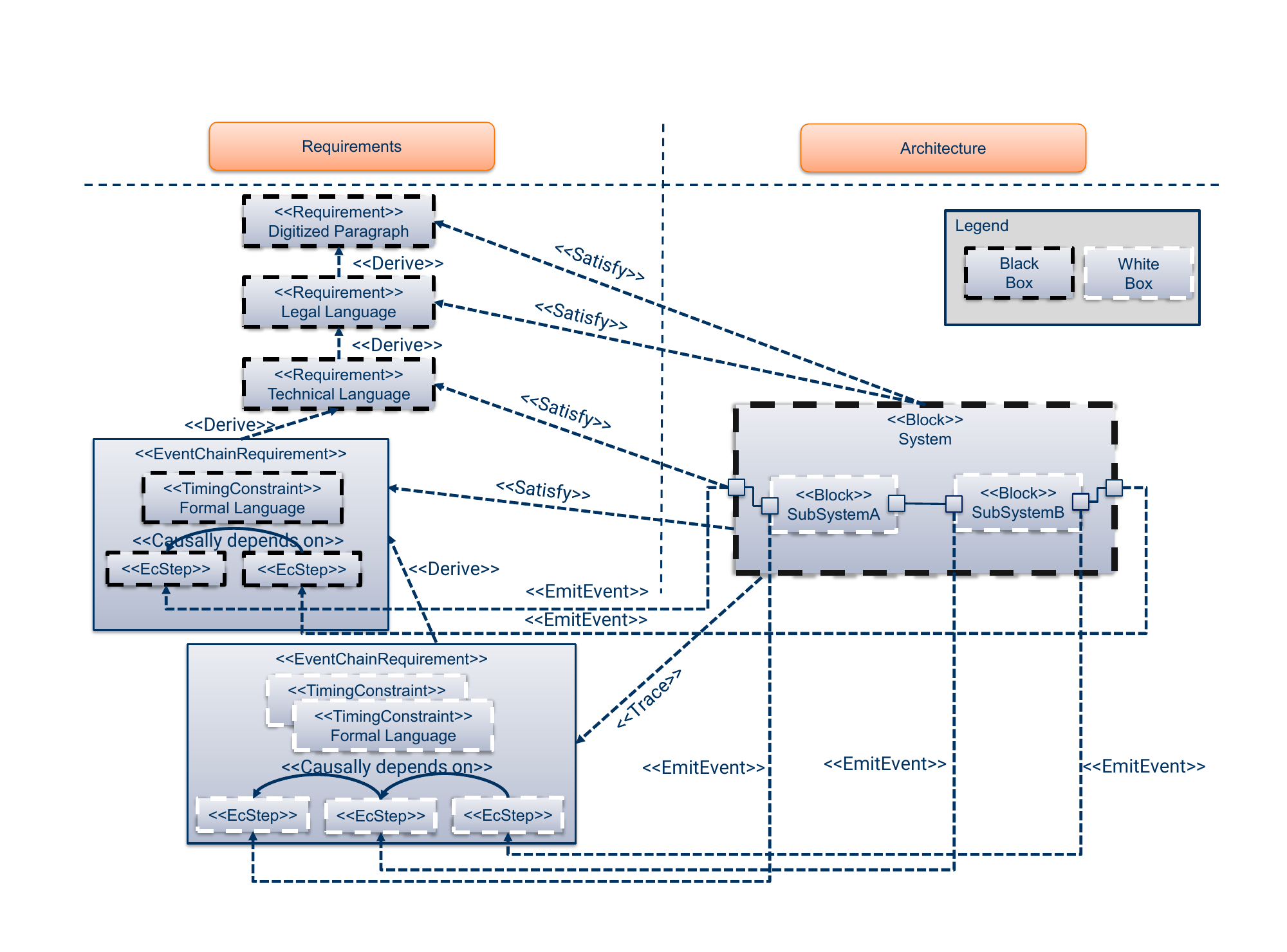}
\caption{Ontology for translating regulations into event-chain models.}
\label{fig:ontology}
\end{figure*}

We propose the ontology depicted in Figure \ref{fig:ontology} to establish architecture description elements and link regulatory requirements to system architecture. 
This structured approach ensures that stakeholder concerns are systematically addressed through dedicated architecture viewpoints and views.

Our methodology follows a multi-stage refinement process, transforming textual legal content into \emph{Event-Chain} models. Each stage narrows the interpretation gap between legal requirements and formal system behavior. During the first steps, the language for the requirements is made more and more formal:

\begin{enumerate}
    \item Starting point is a \textbf{digitized paragraph:} A specific paragraph or clause from a regulation (e.g., UN R152 §5.2.1.1) is identified and extracted. This forms the root of the traceability chain.
    
    \item The digitized paragraph is transformed into \textbf{legal language}, capturing the intent of the regulation in a semi-structured form. These include normative conditions, such as "a warning must be issued at least 0.8 seconds before automatic braking."
    
    \item Legal language is then translated into \textbf{technical language}, which relate to measurable system behavior. For example, "the collision warning function shall be triggered no later than 0.8 seconds before brake command issuance" can be linked to a specific functional signal in the system model.

	%\item Finally, the technical language is transformed into a \textbf{formal language} that can be used for classic requirements (e.g. in a contract based approach \cite{Contracts}) or for Event Chain Models, with Event Chain Requirements and steps.
	\item Finally, the technical-language requirement is transformed into an \textbf{Event-Chain requirement} expressed in formal language, with a corresponding \emph{Event-Chain} Model included as the classifying behavior that defines the expected event structure. 
    
    This modeling step is compatible with the use-case methodology originally proposed by Jacobson, in which use cases serve as the interface between actors and the system to describe externally visible behavior \cite{jacobson_object-oriented_1987}. 
    %In a model-based systems engineering (MBSE) process that follows this philosophy, \emph{Event-Chain} Models can be understood as replacing the behavioral specification of the use cases. 
    %Instead of describing internal function behavior through activity models, the \emph{Event-Chain} Models capture system behavior in terms of observable events and their causal-temporal structure.

\end{enumerate}

%If the requirement is expressed as digitized paragraph or legal language, it can be satisfied by a system as a whole only. If technical or even formal language is used, the expressions can refer to events that occur at ports of the system. This is marked in green in Figure ~\ref{fig:ontology}.

%% file: sec03.02_ec_models.tex
\subsection{Transition from Black-Box to White-Box at the Functional Level}
\label{sec:black-box-2-white-box}

As illustrated in Figure \ref{fig:ontology}, different viewpoints exist on required system behavior and thus on the \emph{Event-Chain Models}. 
The \emph{Black-Box} viewpoint focuses on behavior observable at system boundaries, aiming to formalize stakeholder concerns without anticipating design decisions. 
\emph{Technical Language Requirements} are mapped to \emph{Event-Chain Requirements} of the \emph{Black-Box} viewpoint, with their relation documented via a \code{<<derive>>} dependency. 
Typically, one (sometimes several) \emph{Timing Constraints} capture requirements as formal conditions, each referring to at least two \emph{Event-Chain Steps}.  

In contrast, the \emph{White-Box} viewpoint captures aspects resulting from design decisions—primarily decomposition into subsystems at arbitrary nesting levels. 
For behavioral modeling, \emph{Event-Chain Steps} from the \emph{Black-Box} viewpoint must be refined into one or more \emph{White-Box Event-Chain Steps}. 
This refinement follows the \code{<<causally depends on>>} relationships. 
Thus, a causal dependency that in the \emph{Black-Box} spanned input to output ports can now unfold into a chain of dependencies traversing subsystems and their connections in the \emph{White-Box}. 
Creating a \emph{White-Box} model therefore also yields a corresponding \emph{White-Box Requirement}, again linked to the \emph{Black-Box Model} via \code{<<derive>>}.  

To ensure internal subsystem events remain observable in homologation at the functional level, they must be refined to a technical viewpoint where they become measurable. 
\emph{Event-Chain Engineering} supports this by decomposing a \emph{Black-Box} into a \emph{White-Box Model}: subsystems of the technical viewpoint are anchored by traceability to the functional ones they realize. 
Thus, \emph{Event-Chain Steps} of the functional \emph{White-Box} can be mapped to those of the technical viewpoint.  

At each refinement, whenever end-to-end timing requirements exist, it must be shown that the causal chain of refined \emph{Event-Chain Steps} also satisfies the end-to-end timing constraint. 
Partitioning an end-to-end requirement into one or more timing requirements between intermediate steps is called \emph{budgeting}.  

Different design alternatives—functional or technical—may yield different budgets while fulfilling the same requirement. 
For example, allocating one function to a separate ECU introduces extra communication hops, producing a different budget compared to hosting all on one domain controller.  

Overall, this creates a consistent picture of which subsystem contributes which part to system behavior within a causal chain, and under which requirements. 
The result is documentation that, in line with most regulations, shows not only which observations test correctness but also that the architecture as a whole consistently fulfills the specified requirements.

%% file: sec03.03_analysis.tex
\subsection{Analysis via Event-Chain Models}

Functional decomposition allows the end-to-end behavior of a system to be divided into logically distinct domains such as perception, maneuver planning, and execution. However, this view remains agnostic about \emph{how} these functions are actually realized at runtime. This “how” is captured by the notion of the system’s timing behavior (sometimes also referred to as its \emph{Dynamic Execution Semantics}).

The notion of timing behavior can be expressed at different levels of abstraction. In software architectures, for instance, functions are implemented as runnables, which are mapped to tasks characterized by activation patterns (periodic or event-triggered) and execution priorities. Tasks are then scheduled on compute resources (e.g., cores of an automotive high-performance computer) where they compete according to specific scheduling policies such as fixed-priority preemption.

By contrast, in the communication architecture of a vehicle, interactions between components are realized through messages exchanged on in-vehicle networks. Here, too, contention arises when multiple messages share the same bus, requiring arbitration mechanisms (e.g., priority-based arbitration in CAN, time-sensitive scheduling in Ethernet TSN).

Together, these aspects constitute the system’s \emph{Dynamic Architecture}. Whether such an architecture—across any abstraction level or domain (e.g., ECU scheduling, TSN communication, or a unified system-level timing abstraction)—meets the functional requirements is precisely the question addressed by \emph{Event-Chain Analysis}.

A system-level timing model enables a \emph{shift-left} strategy: by abstracting execution and communication into a unified representation, design flaws can be anticipated long before prototypes exist. Resource sharing and message arbitration are captured through statistically parameterized constructs (e.g., runnable execution times uniformly distributed between $t_{min}$ and $t_{max}$).

\emph{Event-Chain Analysis} can then be performed in complementary ways:

\begin{itemize}
    \item Analytical evaluation e.g., with INCHRON chronVAL\cite{chronVAL2025},

    \item Trace-based validation e.g., with INCHRON chronVIEW\cite{chronVIEW2025},

    \item Simulation e.g., with INCHRON chronSIM\cite{chronSIM2025}, which extends from scheduling-accurate simulation to abstract, system-wide timing analysis.
\end{itemize}

Importantly, \emph{Event-Chain Analysis} is not limited to the system level. \emph{Event-Chain Models} can be decomposed into subsystem-specific chains, allowing detailed timing analysis of individual ECUs, operating systems, or communication domains. Thus, high-fidelity domain tools (e.g., specialized scheduling or TSN analysis environments) naturally complement the broader, system-level perspective.

%% file: sec04_case_study.tex
\section{Case Study: Event-Chain Modeling for Automated Emergency Braking}

This case study focuses on an Automated Emergency Braking (AEB) function in a highway scenario. 
The selected use case involves a vehicle traveling at constant speed in the right lane of a multi-lane highway, approaching a stationary object directly ahead in the same lane. 
Although this represents a strict scenario, it is critical for demonstrating the end-to-end response capability of the AEB system, particularly its ability to detect and respond to static obstacles in high-speed environments.

The system under analysis is responsible for autonomously detecting the hazard and initiating emergency braking when the driver does not react in time. 
The modeled AEB function operates on a simple principle: if an object is present and collision is imminent, the system brakes. 
Note that the model does not include the scenario where an object disappears and braking should stop.
The \emph{Event-Chain} model spans the entire perception–decision–actuation loop and includes the following functions:
\begin{itemize}
    \item \textbf{Data Acquisition ($f_{DA}$):} Sensing the object with an environment sensor such as radar.
    \item \textbf{Object Detection ($f_{OD}$):} Processing of the sensor data to identify potential obstacles.
    \item \textbf{Trajectory Prediction ($f_{TP}$):} Evaluation of own-vehicle dynamics.
    \item \textbf{Collision Assessment ($f_{CA}$):} Determination of when to initiate automated braking based on Time-To-React (TTR).
    \item \textbf{Warning Assessment ($f_{WA}$):} The issuing of a warning to the driver.
    \item \textbf{Brake Control ($f_{BC}$):} Continuously controlling the braking system via the vehicle control unit until the vehicle comes to a complete stop.
\end{itemize}
These functions are allocated across three main components: an independent sensor unit, an ECU handling object detection, trajectory prediction, and collision assessment, and an ECU controlling the braking actuators.

% Three horizontal swim lanes figure (spans both columns)
\begin{figure*}[htbp]
  \centering
  \begin{tikzpicture}[
      swimlane/.style={rectangle, draw, minimum width=17.5cm, minimum height=3cm, align=center},
      process/.style={rectangle, draw, rounded corners, minimum width=2.5cm, minimum height=0.8cm, align=center},
      arrow/.style={->, thick}
  ]
  
  % Define horizontal swim lanes with rotated labels
  \node[swimlane] (lane1) at (0,3.0) {};
  \node[rotate=90] at (-9.2,3.0) {\textbf{Sensor}};

  \node[swimlane] (lane2) at (0,0.0) {};
  \node[rotate=90] at (-9.2,0.0) {\textbf{ECU}};

  \node[swimlane] (lane3) at (0,-3.0) {};
  \node[rotate=90] at (-9.2,-3.0) {\textbf{Actuation ECU}};

  % Define activity diagram styles
  \tikzstyle{start} = [circle, fill=black, minimum size=6pt, inner sep=0pt]
  \tikzstyle{decision} = [diamond, draw, aspect=2, minimum width=3cm, minimum height=1cm, align=center, font=\small]
  \tikzstyle{activity} = [rectangle, draw, rounded corners, minimum width=2.5cm, minimum height=0.8cm, align=center, font=\small]
  \tikzstyle{end} = [circle, draw, thick, minimum size=10pt, inner sep=0pt]

  % Start point
  \node[start] (start) at (-6,4.0) {};
  
  % Sensor activities
  \node[activity] (input) at (-6,3.0) {Data Acquisition};
  \node[activity] (validate) at (2,3.0) {Object Detection};

  % ECU activities
  \node[decision] (warning_check) at (-2.5,0) {Warning Assessment};
  \node[decision] (check) at (6.5,0) {Collision Assessment};
  \node[activity] (analyze) at (2,0) {Trajectory Prediction};
  
  % Actuator ECU activities
  \node[activity] (complete) at (6.5,-3.0) {Brake Control};
  
  % End node (double circle with inner black circle)
  \node[end] (end) at (-2.5,-3.0) {};
  \node[circle, fill=black, minimum size=6pt, inner sep=0pt] at (end) {};

  % Actuator ECU activities
  \node[activity] (warning_issue) at (-7,0.0) {Warning Issuance};
  
  % Activity flow arrows (reworked)
  \draw[arrow] (start) -- (input);
  \draw[arrow] (input) -- (validate);
  \draw[arrow] (validate) -- (analyze);
  \draw[arrow] (analyze) -- (warning_check);
  \draw[arrow] (analyze) -- (check);
  \draw[arrow] (check) -- node[right] {\small Yes} (complete);
  \draw[arrow] (warning_check) -- node[right, yshift=0.2cm] {\small Yes} (warning_issue);
  \draw[arrow] (complete) -- (end);
  \draw[arrow] (warning_issue) |- (end);
  %\draw[arrow] (warning_check) -- (end);

  \end{tikzpicture}
  \caption{Event-chain activity diagram for the AEB case. The chain models the timed flow from perception to actuation: Data Acquisition, Object Detection, Trajectory Prediction, Warning Assessment, Collision Assessment, and Brake Control. Decision nodes split the flow into warning versus braking actions. This formalized sequence defines the end-to-end path on which timing budgets and regulatory checks (e.g., issuing a warning at least 0.8 s before braking) are verified and simulated.}
  \label{fig:swimlanes}
  \end{figure*}
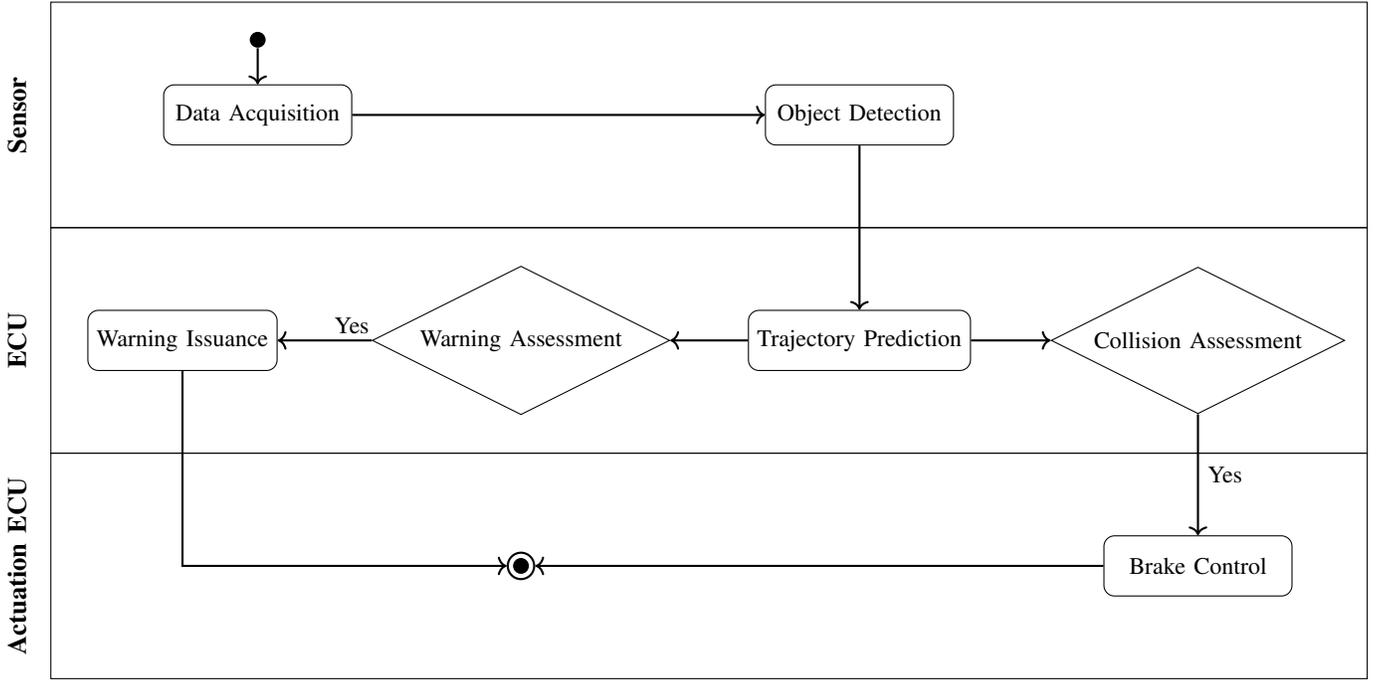

\subsection{Regulatory Requirements}

We evaluate the conformance against the UN Regulation No. 131-2: "Uniform provisions concerning the approval of motor vehicles with regard to the Advanced Emergency Braking Systems (AEBS)".
We will use the following regulatory requirements:
\begin{itemize}
    \item \textbf{LREQ-1:} When the system detects a potential imminent collision, it shall issue a braking demand of at least 4 m/s² to the vehicle's service braking system.
    \item \textbf{LREQ-2:} The collision warning shall be triggered no later than 0.8 seconds before the initiation of emergency braking.
\end{itemize}
These we translate into the following technical requirements:
\begin{itemize}
    \item \textbf{TREQ-1:} The system shall consider a sufficient reaction time to bring the vehicle to a complete stop.
    \item \textbf{TREQ-2:} The system shall issue a warning to the driver no later than 0.8 seconds before the initiation of emergency braking.
\end{itemize}
Note that the digitalization paragraph is omitted here for brevity.

\subsection{Physical Preconditions and System-Level Assumptions}

We model some realistic physical properties that are present in actual systems. Key aspects include:

\begin{itemize}
    \item \textbf{Probabilistic sensor model:} The sensor detection likelihood varies with distance
    \item \textbf{Gradual brake force buildup:} Realistic modeling of brake pressure development over time
    \item \textbf{System response delays:} Accounting for processing and communication latencies
\end{itemize}

The \emph{Event-Chain} is parameterized by reasoning backwards from the final system output (Brake Control) to the initial sensor activity (Data Acquisition), as detailed in the following sections.

\subsubsection{Brake Control}
Brake control continuously manages the braking system until the vehicle comes to a complete stop. The control process is modeled with two phases to reflect actuator behavior:
\begin{enumerate}
    \item A response phase during which deceleration builds up linearly from 0 to $a_{\mathrm{const}}$ with a duration of $t_{\mathrm{response}}$.
    \item A constant-deceleration phase at $a_{\mathrm{const}}$ m/s² with a duration of $t_{\mathrm{const}}$.
\end{enumerate}
During the brake force buildup phase, since the deceleration builds up linearly, the vehicle position follows a cubic function:
\begin{align}
    s_{\mathrm{response}}(v_{\mathrm{initial}}) = v_{\mathrm{initial}} \cdot t_{\mathrm{response}} - \frac{a_{\mathrm{const}} \cdot t_{\mathrm{response}}^3}{6 \cdot t_{\mathrm{response}}}
\end{align}
% Bei 30 m/s und 4 m/s^2 ist die konstante beschleunigung 0.6 s sind es 17.76 m
The time for the constant-deceleration phase $t_{\mathrm{const}}$ depends on the remaining speed $v_{\mathrm{remaining}}$ from the response phase.
\begin{align}
    t_{\mathrm{const}}(v_{\mathrm{initial}}) &= \frac{v_{\mathrm{remaining}}}{a_{\mathrm{const}}} \\
    v_{\mathrm{remaining}} &= v_{\mathrm{initial}} - 0.5 \cdot a_{\mathrm{const}} \cdot t_{\mathrm{response}} \\
\end{align}
The formula depending on the initial speed $v_{\mathrm{initial}}$ and the constant deceleration $a_{\mathrm{const}}$ is:
\begin{align}
    t_{\mathrm{const}}(v_{\mathrm{initial}}) &= \frac{v_{\mathrm{initial}} - 0.5 \cdot a_{\mathrm{const}} \cdot t_{\mathrm{response}}}{a_{\mathrm{const}}}
\end{align}
The distance travelled during the constant-deceleration phase is a quadratic function:
\begin{align}
    s_{\mathrm{const}}(v_{\mathrm{remaining}}) = v_{\mathrm{remaining}} \cdot t_{\mathrm{const}} - \frac{a_{\mathrm{const}} \cdot t_{\mathrm{const}}^2}{2}
\end{align}
The formula depending on the initial speed $v_{\mathrm{initial}}$ and the constant deceleration $a_{\mathrm{const}}$ is:
\begin{align}
    s_{\mathrm{const}}(v_{\mathrm{initial}}) &= (v_{\mathrm{initial}} - 0.5 \cdot a_{\mathrm{const}} \cdot t_{\mathrm{response}}) \cdot t_{\mathrm{const}} \\
    &\quad - \frac{a_{\mathrm{const}} \cdot t_{\mathrm{const}}^2}{2}
\end{align}
% 28.8m * 7.2 s = 207.36 m
% 4 m/s^2 * 7.2 s^2 / 2 = 103.68 m
The total stopping distance is the sum of the distance travelled during the response phase and the distance travelled during the constant-deceleration phase:
\begin{align}
    s_{\mathrm{stop}}(v_{\mathrm{initial}}) = s_{\mathrm{response}}(v_{\mathrm{initial}}) + s_{\mathrm{const}}(v_{\mathrm{initial}})
\end{align}
% 17.76 m + 103.68 m = 121.44 m
The stopping time is the sum of the response time and the constant-deceleration time:
\begin{align}
    t_{\mathrm{stop}} (v_{\mathrm{initial}}) &= t_{\mathrm{response}} + t_{\mathrm{const}}(v_{\mathrm{initial}})
\end{align}
The needed stopping time equals a stopping distance of 

%\begin{figure}
%    \centering
%    \begin{tikzpicture}[xscale=1.8]
        % axes
%        \draw[->] (0,0) -- (4,0) node[below] {$t$ [s]};
%        \draw[->] (0,0) -- (0,4.8) node[left] {$a$ [m/s$^2$]};

        % warning phase: no deceleration (0 to 0.8s) - visually doubled width
%        \draw[thick, red] (0,0) -- (1.6,0);
        
        % braking profile: linear ramp to 4 at 2.2s (1.6 + 0.6s) - visually doubled width, then constant
%        \draw[thick, blue] (1.6,0) -- (2.2,4);
%        \draw[thick, blue] (2.2,4) -- (3.8,4);

        % helper lines
%        \draw[dashed] (1.6,0) -- (1.6,4);
%        \draw[dashed] (2.2,0) -- (2.2,4);
%        \draw[dashed] (0,4) -- (2.2,4);

        % ticks/labels - show actual timing values
%        \node[below] at (1.6,0) {0.8};
%        \node[below] at (2.2,0) {1.4};
%        \node[left] at (0,4) {4};

        % annotations
%        \node at (0.8,0.5) {\small Warning phase};
%        \node at (1.9,2.2) {\small Response phase};
%        \node at (3.0,4.35) {\small Constant deceleration};
%    \end{tikzpicture}
%    \caption{Three-phase braking dynamics used in the timing analysis. A warning phase (0--0.8 s, no deceleration) is followed by a response phase with linear build-up to $4\,\mathrm{m/s^2}$ over $0.6\,\mathrm{s}$, and then a constant-deceleration phase at $4\,\mathrm{m/s^2}$. These phases define the stopping time and distance used in the trigger rules (e.g., for $v=30\,\mathrm{m/s}$: total stopping time $\approx 7.8$ s and distance $\approx 121.44$ m) and provide the reference against which the 0.8 s warning lead-time constraint is checked.}
%    \label{fig:brake_phases}
%\end{figure}

\subsubsection{Trajectory Prediction}
We define the instantaneous \emph{Time-To-React} (TTR) under constant-speed approximation as a function of current object distance \(d_o\) minus the stopping distance \(s_{\mathrm{stop}}(v_{\mathrm{ego}})\) and current ego speed \(v_{\mathrm{ego}}\).
\begin{align}
    \mathrm{TTR}(d_o, v_{\mathrm{ego}}) &= \frac{d_o - s_{\mathrm{stop}}(v_{\mathrm{ego}})}{v_{\mathrm{ego}}}
\end{align}

\subsubsection{Collision Assessment}
We compare the Time-To-React (TTR) with the required stopping time $t_{\mathrm{stop}}$:
\begin{align}
    \mathrm{TTR}(d_o, v_{\mathrm{ego}}) < 0
\end{align}
This function implements \textbf{TREQ-1}, which requires the system to initiate emergency braking no later than when the Time-To-React equals the stopping time.

\subsubsection{Warning Issuance}
Issue a forward-collision warning at least 0.8 s before the braking trigger. 
Equivalently, issue a warning when:
\begin{align}
    \mathrm{TTR}(d_o, v_{\mathrm{ego}}) < 0.8\,\mathrm{s}
\end{align}
This function is responsible for implementing \textbf{TREQ-2}, which requires the system to issue a warning to the driver no later than 0.8 seconds before the initiation of emergency braking.

\subsubsection{Object Detection}
Let's define the guaranteed object distance $r_{g}$ as the distance at which the sensor is guaranteed to detect an object.
Realistically, this is the minimum distance at which the sensor can detect an object.
However, we assume that the sensor can detect objects up to a maximum distance $r_{max}$ and that the detection likelihood decreases exponentially with distance.
In \cite{Wang.2010} this is called truncated sensor model. 
%Figure~\ref{fig:sensor_function} illustrates this sensor model, showing the detection likelihood as a function of distance to the object.
\begin{align}
    f_{ES} &: \mathbb{R} \rightarrow \{0,1\} \\
    f_{ES}(d_o) &=
    \begin{cases}
        1 & d_o \leq r_g \\
        e^{-\frac{d_o - r_g}{r_{max} - r_g}} & r_g < d_o \leq r_{max} \\
        0 & d_o > r_{max}
    \end{cases}
\end{align}

%\begin{figure}
%    \centering
%    \includegraphics[width=0.5\textwidth]{../../05_Simulation/sensor_function_plot.pdf}
%    \caption{Phenomenological sensor detection model as a function of distance. A guaranteed detection zone (e.g., 0--100 m) is followed by an exponential decay region (e.g., 100--140 m) where probability drops from 1.0 to 0.2. Parameterized by the guaranteed range $r_g$ and maximum range $r_{max}$, this function determines whether an object is detected early enough to satisfy the 0.8 s warning requirement (e.g., at $\approx$145 m in the case study) and directly feeds the Monte Carlo evaluation of compliance probability.}
%    \label{fig:sensor_function}
%\end{figure}

%Since the issuing of a warning depends on object distance outside of the guaranteed object distance $r_{g}$, we need to simulate the whole \emph{Event-Chain} model to assess the probability of the warning being issued.

\subsection{Event-Chain Model}

\begin{figure}[h!]
    \centering
    \includegraphics[width=0.5\textwidth]{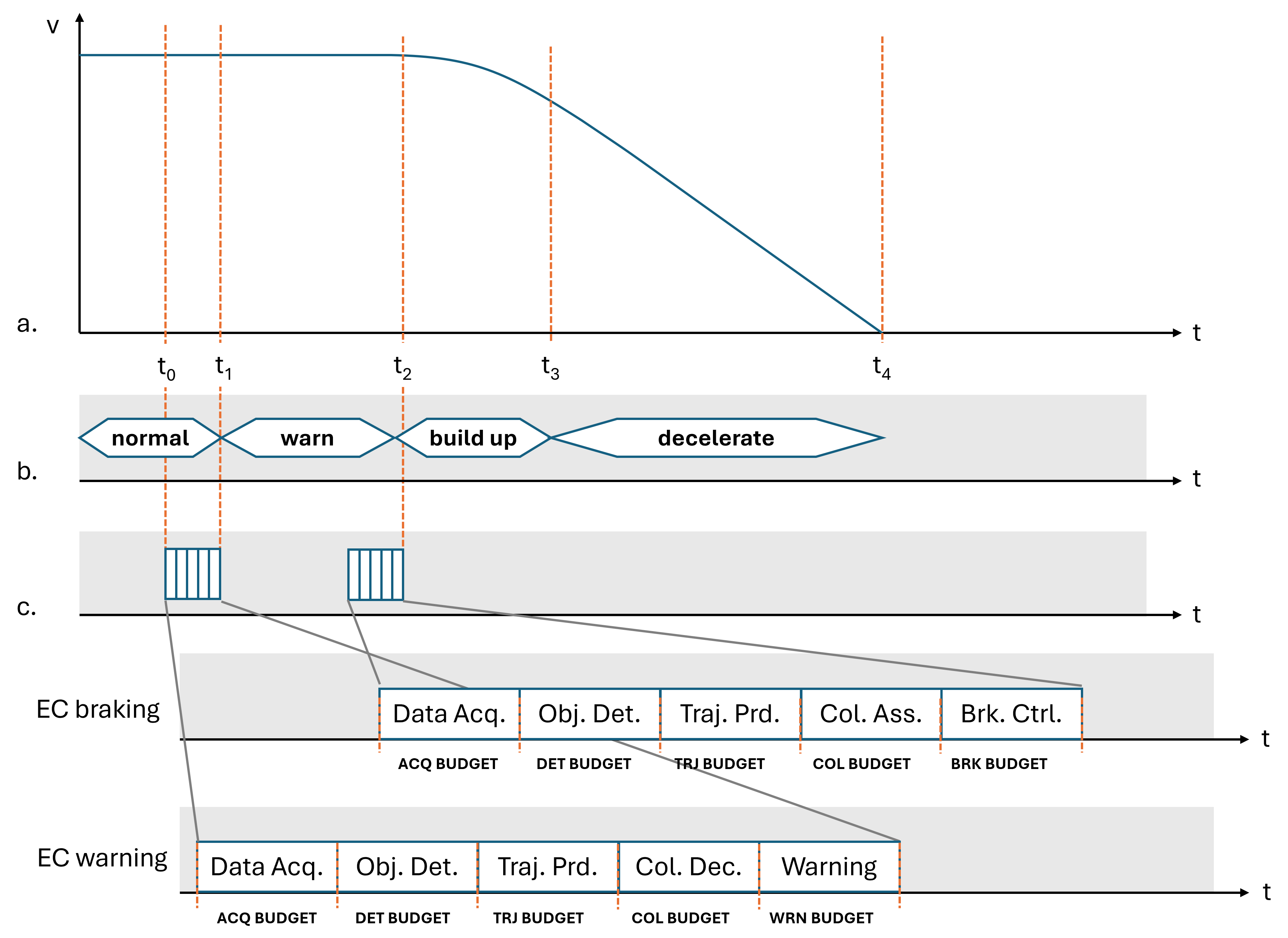}
\caption{Braking scenario. a. Speed–time diagram of the ego vehicle, b. \emph{Black-box} \emph{Event-Chain}, c. \emph{White-box} \emph{Event-Chains}: \emph{EC Braking} and \emph{EC Warning}}
\label{fig:scenario}
\end{figure}

% The event-chain model is shown in Figure~\ref{fig:swimlanes}.

Following the methodology outlined in Section~\ref{sec:black-box-2-white-box}, we first construct a \emph{Black-Box Event-Chain} that serves as a basis for expressing the formalized requirements \textbf{ECREQ-1} and \textbf{ECREQ-2} in the form of timing constraints, which are derived from \textbf{TREQ-1} and \textbf{TREQ-2}. In the next step, this \emph{Black-Box Event-Chain} is refined into a \emph{White-Box Event-Chain}. Based on the available system design—including both hardware and software components—additional \emph{White-Box} constraints are defined, which are intended to be monitored within the system. The simulation of the system design enables concrete verification of both \emph{Black-Box} and \emph{White-Box} constraints.

Figure \ref{fig:scenario} illustrates this three-stage refinement process: the scenario description, the \emph{Black-Box} \emph{Event-Chain}, and the \emph{White-Box} \emph{Event-Chain} with associated timing requirements.

According to requirements \textbf{TREQ-1} and \textbf{TREQ-2}, the following sequence of events must be observable at the system boundary\footnote{The previously derived requirements could, in principle, be represented as two distinct \emph{Event-Chains} (one for warning and one for braking). However, since both rely on the same sequence of sensor input, object detection, and trajectory prediction, they are modeled together as a single Event Chain.}:

\begin{itemize}
\item $t_0$: An obstacle enters the sensor’s detection range.
\item $t_1$: The driver warning is issued.
\item $t_2$: The vehicle transitions from cruising to deceleration—observable through brake pressure buildup.
\item $t_3$: The vehicle decelerates with at least $a_{\mathrm{const}}$
\item $t_4$: The vehicle comes to a complete stop in front of the obstacle.
\end{itemize}

Based on this event sequence, the following formal timing requirements are defined:

\begin{itemize}
    \item \textbf{ECREQ-1.1:} $\frac{v(t_4)-v(t_2)}{t_4-t_2} \ge a_{\mathrm{const}}$ 
    \item \textbf{ECREQ-1.2:} $t_4-t_2 \ge \mathrm{TTR}(v_0)$
    \item \textbf{ECREQ-2:} $t_2-t_1 \ge 800 ms$
\end{itemize}
%Since \textbf{ECREQ-1.2} depends on the initial vehicle speed $v_0$, a Monte Carlo simulation can be used to evaluate the constraint across a distribution of $v_0$ values.

The \emph{Black-Box} \emph{Event-Chain} described above is to be realized by the causally consistent interaction of the following functional elements: \textbf{Data Acquisition ($f_{DA}$)}, \textbf{Object Detection ($f_{OD}$)}, \textbf{Trajectory Prediction ($f_{TP}$)}, \textbf{Collision Assessment ($f_{CA}$)}, \textbf{Warning Assessment ($f_{WA}$)} and \textbf{Brake Control ($f_{BC}$)}.

The causal dependencies between these functions—omitted in Figure~\ref{fig:scenario}—are represented within the \emph{Event-Chain} model using \code{<<causally depends on>>} relationships.

Based on physically and technically realistic assumptions, the following \emph{White-Box} timing requirements are defined:

\begin{itemize}
    \item \textbf{ACQ BUDGET} $t_{acq} \leq 450 ms$  
    \item \textbf{DET BUDGET} $t_{det} \leq 10 ms$
    \item \textbf{TRJ BUDGET} $t_{trj} \leq 30 ms$
    \item \textbf{COL BUDGET} $t_{col} \leq 10 ms$
    \item \textbf{WRN BUDGET} $t_{wrn} \leq 800 ms$
\end{itemize}

Due to the probabilistic nature of the sensor, an additional requirement is defined for detection quality at time $t_{det}$:

\begin{itemize}
    \item \textbf{DET ERROR RATE} $e_{det, t_{det}} \leq 0.01$ 
\end{itemize}

\subsection{Event-Chain Simulation}
\begin{figure}[h]
    \centering
    \includegraphics[width=0.5\textwidth]{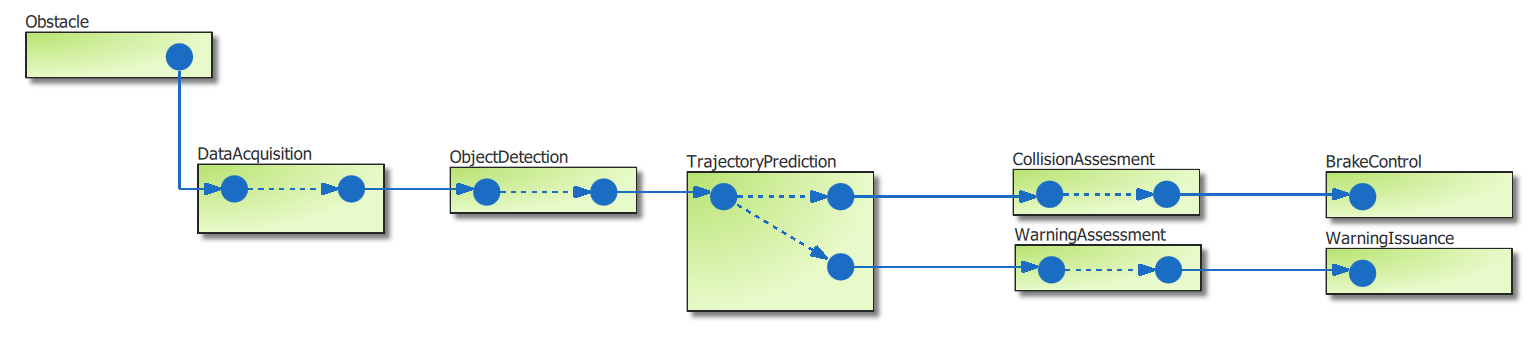}
\caption{Event-chain for the AEB case in chronSUITE}
\label{fig:ec_chronsuite}
\end{figure}

In order to verify that the requirements derived from regulations are fulfilled, the \emph{Event-Chain} model has to be transformed into a simulation model. The structure is shown in Figure~\ref{fig:ec_chronsuite}. The blocks correspond to the architecture that shall satisfy the requirements derived from regulatory. Additionally some stimuli have to be added to make the model executable. The simulation model also contains requirements that are evaluated on the simulation results.

%\begin{weak}
%We also have to define timing constraints on the modeled Event-Chains. \textbf{ECREQ-1} is derived from \textbf{TREQ-1} and specifies the required time for braking. It consists of the time required for deceleration itself which is 7.8s at $30\,\mathrm{m/s}$ and time 0.3s for processing in the E/E system. In \emph{Black-Box View} the constraint of 8.1s has to be satisfied by the system as a whole. In \emph{White-Box View} it can be broken down into 0.1s from detection to decision on braking, 0.2s from decision to start of actual braking and 7.8s for the braking process itself. \textbf{ECREQ-2} is derived from \textbf{TREQ-2} which is 0.8s between start of braking and warning.
%\end{weak}

%The event-chain from figure \ref{fig:swimlanes} can be transferred to a simulation model. This is shown in figure~\ref{fig:ec_chronsuite}.

This makes it possible to simulate the dynamic system behavior and check it against the technical requirements TREQ-1 and TREQ-2.
We use chronSIM\cite{chronSIM2025} for this purpose, which is able to account for the probabilistic sensor model by using a Monte Carlo simulation. 
The tool's flexibility is demonstrated through its ability to implement proprietary functions via embedded C-code, allowing for the integration of domain-specific models such as the truncated sensor model used in this case study.

Summing up the budgets before actual braking ($t_{acq}, t_{det}, t_{trj}, t_{col}$) gives 500ms. This means at 30 m/s we need an additional detection distance of 15m for the sensor. Considering this, we parameterize the simulation model with the following parameters:
\begin{itemize}
    \item Initial speed $v_{\mathrm{initial}} = 30\,\mathrm{m/s}$ 
    \item Constant deceleration $a_{\mathrm{const}} = 4\,\mathrm{m/s^2}$
    \item Response time $t_{\mathrm{response}} = 0.6\,\mathrm{s}$
    %\item Guaranteed object distance $r_{g} = 121.44\,\mathrm{m}$
    %\item Maximum object distance $r_{max} = 145.44\,\mathrm{m}$
    \item Guaranteed object distance $r_{g} = 136.44\,\mathrm{m}$
    \item Maximum object distance $r_{max} = 160.44\,\mathrm{m}$
    \item Sensor sampling frequency $f_{\mathrm{sensor}} = 10\,\mathrm{Hz}$
	\item Trajectory Prediction frequency $f_{\mathrm{trj}} = 25\,\mathrm{Hz}$
	\item Brake Control frequency $f_{\mathrm{brake}} = 100\,\mathrm{Hz}$
\end{itemize}

\subsubsection{Simulation Results}
Figure~\ref{fig:result_combined} illustrates the resulting distribution of warning lead times under the phenomenological model.
The ego vehicle is braking early enough to avoid a collision.
Therefore, the \emph{Event-Chain} model is compliant with the regulatory requirements TREQ-1 and with that with LREQ-1.

In this configuration the 0.8 s requirement on warning is violated in $\approx$ 20\% of the evaluations; hence the configuration does not meet the regulatory requirements TREQ-2.

The parameterization of the \emph{Event-Chain} models needs to be adjusted to meet the regulatory requirements TREQ-2.

\subsubsection{Optimization}
% Case Study could show with changing certain parameters the overall latency is met. Analog to \cite{Heckmann.2021}.
%Verification of end-to-end latency requirements. \cite{Dube.2016}

The \emph{Event-Chain} model can be optimized to meet the regulatory requirement of meeting the warning lead time of 0.8 seconds by adjusting the following levers:

\begin{itemize}
    \item The guaranteed object distance $r_{g}$
    \item The maximum object distance $r_{max}$
    \item Frequency of the sensor sampling
\end{itemize}

\begin{figure}
    \centering
    \includegraphics[width=0.5\textwidth]{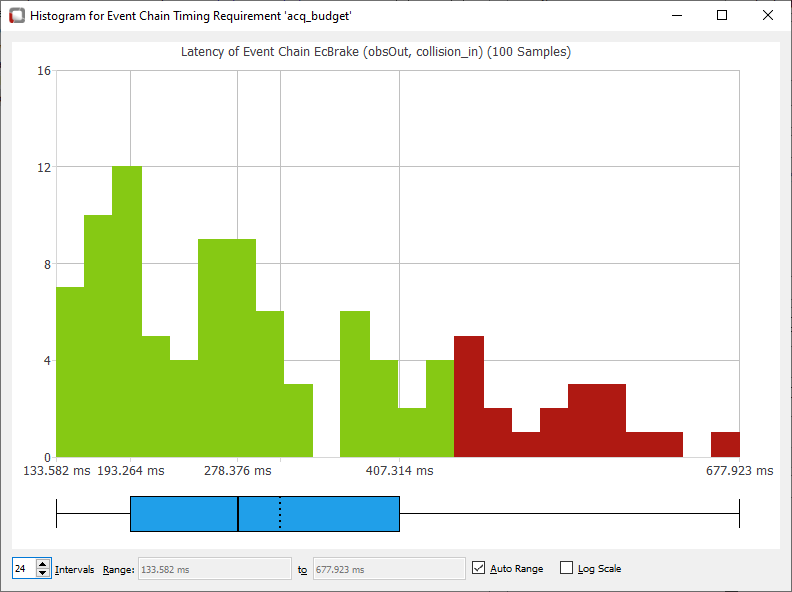}
\caption{Histogram of requirement on budget $t_{acq}$}
\label{fig:result_acq}
\end{figure}

Detailed analysis of the simulation results shows that the violation of budget $t_{acq}$ is the root cause for the requirements violations. This is shown in the histogram in Figure~\ref{fig:result_acq}.

\begin{figure}[h!]
  \centering
  \begin{subfigure}[b]{0.24\textwidth}
    \centering
    \includegraphics[width=\textwidth]{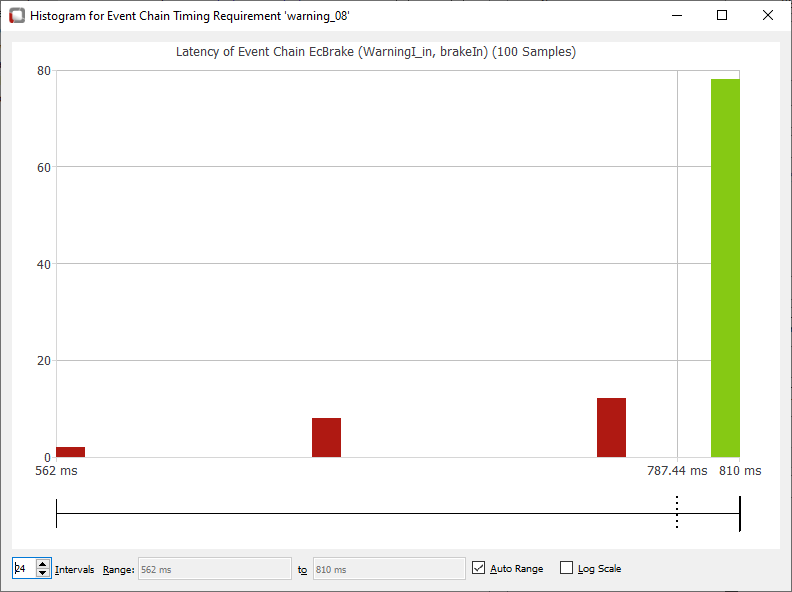}
    \caption{Baseline (10 Hz sampling): $\approx$ 20\% violations 0.8 s requirement on warning.}
    \label{fig:result}
  \end{subfigure}
  \hfill
  \begin{subfigure}[b]{0.24\textwidth}
    \centering
    \includegraphics[width=\textwidth]{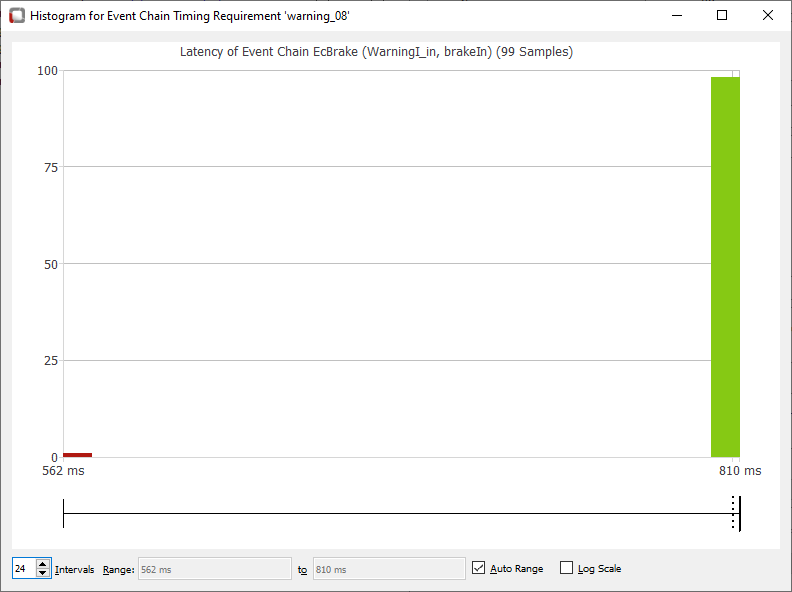}
    \caption{Optimized (20 Hz sampling): only $\approx$ 1\% violations, satisfying the requirement.}
    \label{fig:result_optimized}
  \end{subfigure}
  \caption{Monte Carlo evaluation of warning lead times relative to brake trigger. (a) Baseline configuration (10 Hz sampling) underperforms with 20\% violation sof the 0.8 s requirement. (b) Optimized configuration (20 Hz sampling) makes 99\% meet the 0.8 s threshold, demonstrating restored compliance under otherwise unchanged assumptions.}
  \label{fig:result_combined}
\end{figure}

In this case study we demonstrate the optimization of the \emph{Event-Chain} by changing the frequency of the sensor sampling in order to meet the budget $t_{acq}$.
We can see that by increasing the frequency of the sensor sampling from 10 Hz to 20 Hz, the average warning lead time is increased to 0.8 seconds.
Therefore, the system is compliant with the regulatory requirements.
While increasing sampling frequency improves warning lead time, it introduces practical trade-offs: higher sensor and ECU duty cycles increase power consumption and thermal load; higher data rates raise bus utilization (e.g., CAN/FlexRay/Ethernet) and buffering needs; tighter scheduling may reduce timing slack for co-located functions. 
In practice, engineers balance these effects against alternatives (e.g., improving detection probability, extending sensor range, or optimizing processing pipelines) to meet requirements within platform constraints.
Figure~\ref{fig:result_combined} shows the histogram of the warning being issued after optimization.
We see that on average the warning is issued 0.8 seconds before the braking is initiated.
Therefore, the system is compliant with the regulatory requirements.

